\documentclass[onecolumn,unpublished,a4paper]{quantumarticle}
\usepackage[backend=biber,style=alphabetic,maxbibnames=999]{biblatex}
\usepackage{amsmath}
\usepackage{amssymb}
\usepackage{amsthm}
\usepackage{amsfonts}
\usepackage[caption=false]{subfig}
\usepackage[colorlinks]{hyperref}
\usepackage[all]{hypcap}
\usepackage{tikz}
\usepackage{relsize}
\usepackage{color,soul}
\usepackage[utf8]{inputenc}
\usepackage{capt-of}
\usepackage{mathtools}
\usepackage{float}
\usepackage[section]{placeins}
\usepackage{listings}
\usepackage[T1]{fontenc}  
\usetikzlibrary{decorations.pathreplacing}
\usepackage{braket}

\DeclareFixedFont{\ttb}{T1}{txtt}{bx}{n}{5}
\DeclareFixedFont{\ttm}{T1}{txtt}{m}{n}{5}
\definecolor{deepblue}{rgb}{0,0,0.5}
\definecolor{deepred}{rgb}{0.6,0,0}
\definecolor{deepgreen}{rgb}{0,0.5,0}
\newcommand\cppstyle{\lstset{
language=C++,
basicstyle=\ttm,
otherkeywords={uint8_t, __m256i, size_t, ASSERT_TRUE, EXPECT_TRUE, TEST, BENCHMARK},
keywordstyle=\ttb\color{deepblue},
emphstyle=\ttb\color{deepblue},
stringstyle=\color{deepgreen},
commentstyle=\fontfamily{txtt}\selectfont\color{gray},
showstringspaces=false,
literate={*}{{\char42}}1
         {-}{{\char45}}1
}}
\lstnewenvironment{cpp}[1][]
{\cppstyle\lstset{#1}}{}

\newcommand\pythonstyle{\lstset{
language=python,
basicstyle=\ttm,
morekeywords={assert,as,echo},
keywordstyle=\ttb\color{deepblue},
emphstyle=\ttb\color{deepblue},
stringstyle=\color{deepgreen},
commentstyle=\fontfamily{txtt}\selectfont\color{gray},
showstringspaces=false,
literate={*}{{\char42}}1
         {-}{{\char45}}1
}}
\lstnewenvironment{python}[1][]
{\pythonstyle\lstset{#1}}{}

\lstdefinestyle{stimcircuit}{
    language=python,
    basicstyle=\fontsize{4}{4}\selectfont\ttfamily,
    upquote=true,
    stepnumber=1,
    numbersep=8pt,
    showstringspaces=false,
    breaklines=true,
    frame=single,
    aboveskip=1.5em,
    belowskip=1.5em,
    commentstyle=\color{gray},
    classoffset=1,
    morekeywords={DETECTOR,OBSERVABLE_INCLUDE,rec},
    keywordstyle=\color{deepgreen},
    classoffset=2,
    morekeywords={H,R,MPP,M},
    keywordstyle=\ttb\color{deepblue},
    classoffset=3,
    morekeywords={X_ERROR,DEPOLARIZE2,DEPOLARIZE1},
    keywordstyle=\color{red},
    classoffset=4,
    morekeywords={TICK,SHIFT_COORDS,QUBIT_COORDS},
    keywordstyle=\color{gray}
}

\renewcommand\thesection{\arabic{section}}

\theoremstyle{definition}

\theoremstyle{definition}

\theoremstyle{definition}

\newcommand{\eq}[1]{\hyperref[eq:#1]{Equation~\ref*{eq:#1}}}
\renewcommand{\sec}[1]{\hyperref[sec:#1]{Section~\ref*{sec:#1}}}
\DeclareRobustCommand{\app}[1]{\hyperref[app:#1]{Appendix~\ref*{app:#1}}}
\newcommand{\fig}[1]{\hyperref[fig:#1]{Figure~\ref*{fig:#1}}}
\newcommand{\tbl}[1]{\hyperref[tbl:#1]{Table~\ref*{tbl:#1}}}
\newcommand{\theoremref}[1]{\hyperref[theorem:#1]{Theorem~\ref*{theorem:#1}}}
\newcommand{\definitionref}[1]{\hyperref[definition:#1]{Definition~\ref*{definition:#1}}}

\begin{document}
\title{Stability Experiments: The Overlooked Dual of Memory Experiments}

\date{\today}
\author{Craig Gidney}
\email{craig.gidney@gmail.com}
\affiliation{Google Quantum AI, Santa Barbara, California 93117, USA}

\begin{abstract}
Topological quantum computations are built on a foundation of two basic tasks:
preserving logical observables through time
and moving logical observables through space.
Memory experiments, which check how well logical observables are preserved through time, are a well established benchmark.
Strangely, there is no corresponding well established benchmark for moving logical observables through space.
This paper tries to fill that gap with ``stability experiments", which check how well a quantum error correction system can determine the product of a large region of stabilizers.
Stability experiments achieve this by testing on a region that is locally a normal code but globally has a known product of stabilizers.
\end{abstract}

\maketitle

\emph{The source code that was written, the exact noisy circuits that were sampled, and the statistics that were collected as part of this paper are available at \href{https://doi.org/10.5281/zenodo.6859486}{doi.org/10.5281/zenodo.6859486}~\cite{craig_gidney_2022_6859486}.}

\section{Introduction}
\label{sec:introduction}

\begin{figure}[b!]
    \centering
    \resizebox{\linewidth}{!}{
        \includegraphics{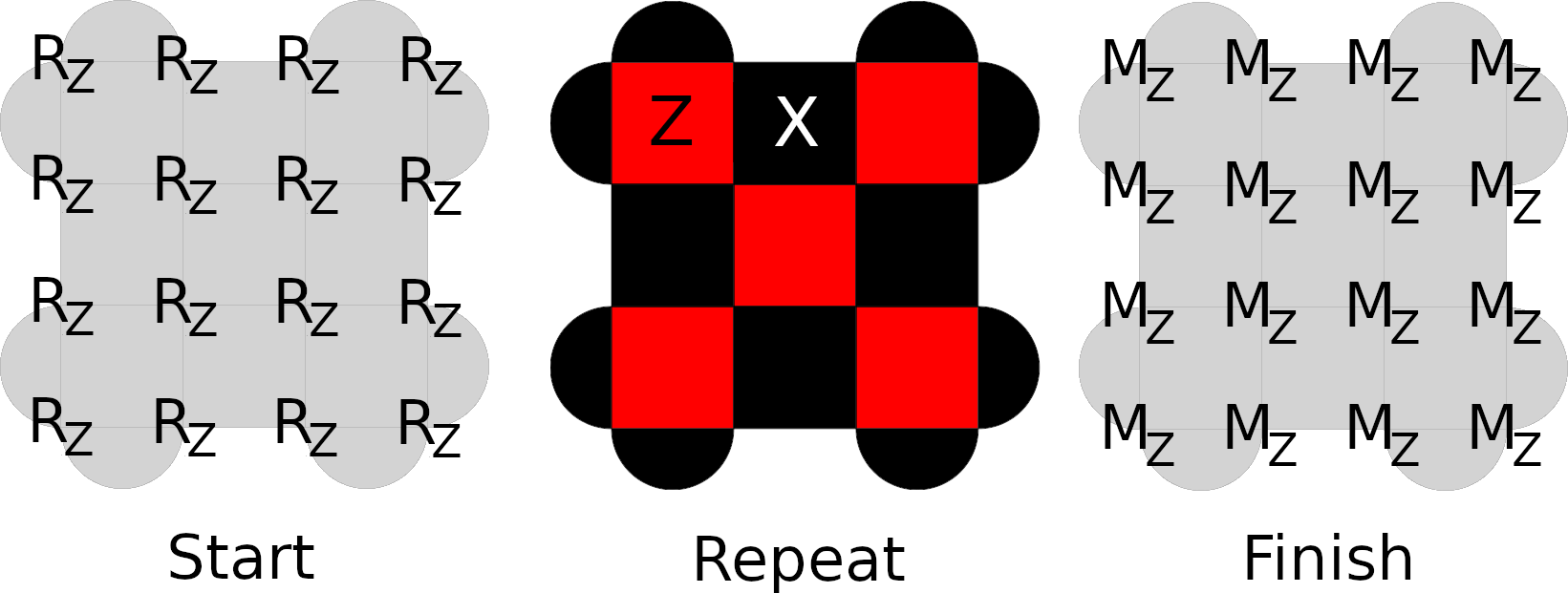}
    }
    \caption{
        A 4x4 X type surface code stability experiment.
        The experiment begins by initializing the 16 data qubits into the $|0\rangle$ state.
        Then the stabilizers (of the 4x4 surface code patch with X type boundaries all the way around) are repeatedly measured.
        The 16 data qubits are then measured in the Z basis.
        Finally, after error correction, it's verified that the combined product of all twelve X stabilizers was $+1$.
        If the product was $-1$, a logical error occurred.
    }
    \label{fig:4x4_X}
\end{figure}

In a quantum error correcting code, such as the surface code, there are many different but equivalent logical observables that you can track.
For example, in a surface code patch with Z type boundaries at the top and bottom, the Z logical observable could be a vertical path along the left hand side of the patch, or along the right hand side of the patch, or many other possibilities.
Each choice of observable has an associated ``byproduct operator"~\cite{fowler2012surfacecodereview}, which will be either $+1$ or $-1$, indicating whether the observable is currently negated compared to what the overlying quantum algorithm expects.
A key distinction between the various choices of which equivalent observable to track is that the different choices can have different byproduct operators.
Moving from one choice of logical observable to another is a book-keeping operation, where the relationship between the byproduct operators is determined by the measurement results from the stabilizers that separate the observables.
So, ultimately, moving a logical observable through space comes down to correctly multiplying the contributions of many stabilizer measurements into its byproduct operator.

For example, consider a system with a logical observable $X_L = +X_1 X_2 X_3$ and a measured stabilizer observable $X_S = +X_1 X_2 X_4 X_5$.
Suppose the stabilizer measurement result is $-1$, after error correction, meaning you are confident that $-X_S = +1$.
From this information you can derive $X_L = X_L \cdot +1 = X_L \cdot -X_S = -X_3X_4 X_5$.
In other words, $X_S$ tells you how to express the logical observable $X_L$ in terms of qubits 3, 4, and 5 instead of qubits 1, 2, and 3.
It allows you to move the logical observable from being supported by qubits 1, 2, and 3 (with byproduct operator $+1$) to being supported by qubits 3, 4, and 5 (with byproduct operator $-1$).

In realistic scenarios, with big code distances or long routing distances, moving a logical observable will involve multiplying hundreds or even millions of stabilizers into the observable's byproduct operator.
If \emph{any one} (or three, or five, or etc) of the measurements of those stabilizers is wrong, the sign of the moved logical observable will be wrong.
This is a logical error; a disastrous situation where the overlying algorithm being executed by the quantum computer will silently produce bad results.

The relevance of computing large products of stabilizers to fault tolerant quantum computation is well known in the quantum error correction field \cite{Raus07d,horsman2012latticesurgery,chamberland2022building,chamberland2022universal,chamberland2022circuit}.
Moving logical observables requires multiplying many stabilizers together, and it's impossible to do any computation if you just leave everything in the same place forever.
Therefore, being able to reliably compute huge products of stabilizers is extremely important.
Given these facts, it's strange that there's no well established experiment for directly verifying the ability to compute large stabilizer products (analogous to how memory experiments are a well established benchmark for directly verifying the ability to preserve a qubit through time \cite{chen2021exponential,ryan2021realization,zhao2021realizing,krinner2021realizing,andersen2020repeated}).
The goal of the experiment type proposed by this paper, ``stability experiments", is to fill this gap.

\begin{figure}
    \centering
    \resizebox{\linewidth}{!}{
        \includegraphics{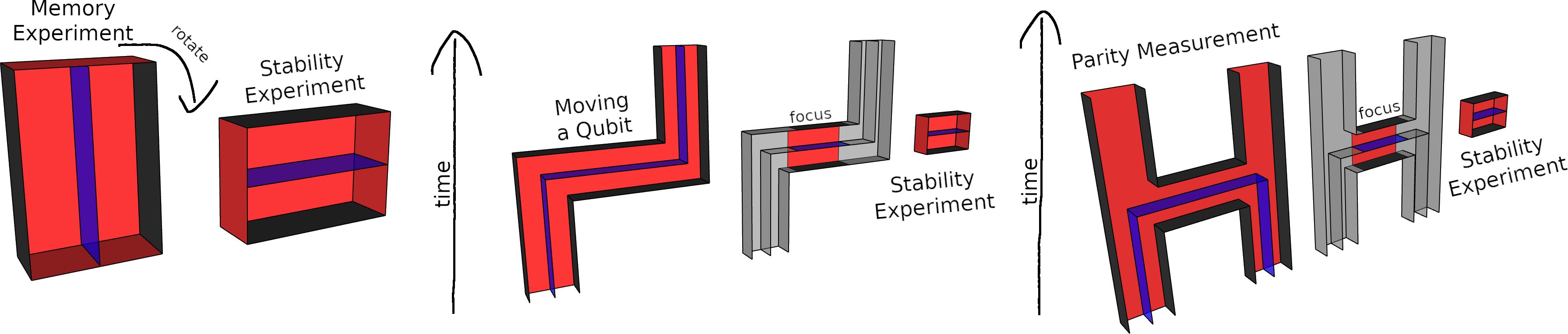}
    }
    \caption{
        Various ways to derive the stability experiment from topological spacetime diagrams.
        (Left) the stability experiment is a spacetime rotation of the memory experiment.
        (Middle) the stability experiment verifies a spacelike part of moving one of the observables of a logical qubit.
        (Right) the stability experiment verifies a spacelike part of computing a parity measurement's result.
        Red surfaces are Z type boundaries.
        Black surfaces are X type boundaries.
        Blue surfaces are valid parity sheets.
        Front faces are omitted so interior details can be seen.
    }
    \label{fig:topological}
\end{figure}

At a high level, stability experiments are actually quite similar to memory experiments (see \fig{topological}).
Memory experiments work because they setup a situation with a global invariant across time, and then check that invariant.
The invariant is that the state measured at the end of time is supposed to match the state prepared at the start of time.
This makes memory experiments somewhat degenerate.
The measurement result is known ahead of time, so there's no algorithmic need to do all those expensive quantum operations at runtime.
In a large quantum computation, you would want to optimize away anything that looked like a memory experiment.

Stability experiments also work by creating and verifying a global invariant.
The main difference is that, instead of using a global invariant across time, stability experiments setup a situation with a global invariant across space.
Specifically, it will be the case during a stability experiment that a region of stabilizers has a product whose correct value is known ahead of time.
This makes stability experiments somewhat degenerate, like memory experiments, in that in practice in a large quantum computation you would want to optimize away anything that looked like a stability experiment.
Still, by avoiding the urge to delete the degeneracy, you can compare the product computed at runtime with the known correct value.
This allows you to determine how good your error correction system is at quickly determining these products of regions of stabilizers.

There are several reasons to be interested in the outcome of stability experiments.
For example, the stability experiment can be used to determine how many rounds are needed to achieve a desired level of certainty that a logical qubit was moved correctly.
More generally, the stability experiment can be used to quantify whether the ``timelike code distance" (the number of times stabilizer measurements are repeated) needs to be smaller or larger than the ``spacelike code distance" (the diameter of surface code patches).
Usually it's assumed that these numbers are the same, but there's no strict reason that they have to be.
A more abstract justification for being interested in the stability experiment is given in \fig{topological}: the stability experiment is hiding in the topological spacetime diagrams of common quantum computations.

A final reason to be interested in the stability experiment is that, because its code distance depends on how long it runs instead of how many qubits it covers, it allows experimental access to arbitrarily high code distances.
Additionally, as I'll show with simulation results, at small patch sizes the stability experiment is easier than memory experiments.
A quantum computer with a noise level above the threshold of the surface code (meaning making surface code patches bigger increases the logical error rate rather than decreasing it) can still be below the threshold of small fixed-patch-size stability experiments (meaning that increasing the number of rounds decreases the logical error rate rather than increasing it).
This suggests that a quantum computer near (but not yet below) the threshold of the surface code may still be able to demonstrate strong exponential suppression of a relevant logical error mechanism; to demonstrate progress towards the ultimate goal of strong exponential suppression of \emph{all} logical error mechanisms at full scale.

\section{The Stability Experiment}

Consider a surface code patch that has the same boundary type all the way around the patch.
For example, \fig{4x4_X} shows a 4x4 patch surrounded by an X type boundary.
This patch cannot store a logical qubit, because there is no place to terminate a Z observable.
However, this patch does have a known product of stabilizers.
Notice that each data qubit participates in exactly two X type stabilizers.
This means that, if you compute the product of all of the X type stabilizers, each data qubit $q$ will contribute two $X_q$ terms to the product and these terms will all cancel out.
So the product of all the X stabilizers must be $\prod_{q \in \text{data}} (X_q)^2 = +1$; a constant known ahead of time.

Because we're considering a situation with noise, the stabilizer measurement results won't always satisfy this global stabilizer product invariant.
If one of the X stabilizer measurements is flipped, the product of all the X stabilizer measurements will also be flipped.
It will be $-1$ instead of $+1$.
Counter-intuitively, performing local error correction doesn't ensure the global product is correct.

\begin{figure}[ht!]
    \centering
    \resizebox{\linewidth}{!}{
        \includegraphics{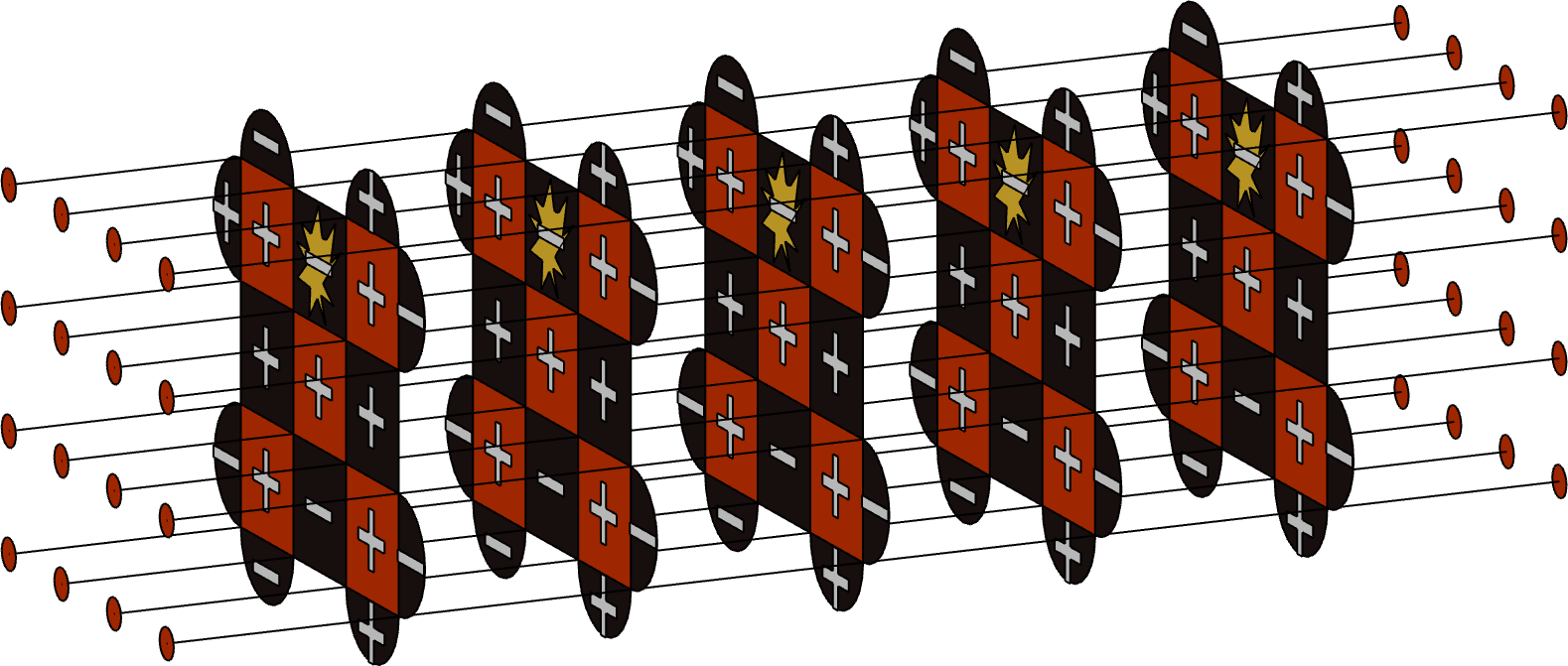}
    }
    \caption{
        Example of an undetected logical error in a stability experiment.
        At the start of time (on the left), the data qubits are initialized in the Z basis.
        At the end of time (on the right), the data qubits are measured in the Z basis.
        In between, the stabilizers of the code are measured five times.
        The Z basis stabilizers are deterministically +1 due to the initialization, whereas the X basis stabilizers project randomly to either -1 or +1.
        Each round, each stabilizer is annotated with its reported measurement (``+'' for +1 and ``-'' for -1).
        Every time the stabilizers are measured, a measurement error occurs on the top center X basis stabilizer (indicated by yellow splotches).
        Its actual value is +1, but it keeps being reported as -1.
        Because the measurement errors occur consistently on the same stabilizer, and because the data initialization and data measurement occur in the Z basis, no detection events are produced.
        The decoder will predict that no corrections are needed, because locally it appears that nothing is going wrong (there are no detection events).
        But, globally, the product of all X stabilizers is -1 instead of +1.
        The decoder predicted incorrectly.
        A logical error occurred.
    }
    \label{fig:logical_error}
\end{figure}

The way errors are corrected in the surface code is by looking for ``detection events"; for stabilizer measurements not matching some local invariant.
For example, a $Z^{\otimes 4}$ measurement returning $-1$ when the data qubits were just initialized into $|0\rangle$ is a detection event, because $|0\rangle^{\otimes 4}$ is in the $+1$ eigenbasis of $Z^{\otimes 4}$ not the $-1$ eigenbasis.
Another example of a detection event is a stabilizer measurement returning a different result than it did in the previous round, since these measurements should be staying stable over time.
If there are no detection events, the error correction system will assume no errors have occurred and make no corrections.
To show that the stability experiment can experience a logical error, it's sufficient to give an example where there are no detection events but the global product of stabilizers is wrong.
Such an example is shown in \fig{logical_error}.
The key feature is a chain of measurement errors stretching from the start of the experiment to the end of the experiment.

Note that the logical error shown in \fig{logical_error} is detectable if the initialization and measurement of the data qubits is done in the same basis as the type of the spatial boundaries.
For a logical error to exist, the time boundaries (the data initialization and the data measurement) must not be the same type as the spatial boundaries.
This is a general requirement of the stability experiment.
From a topological perspective, this is because all error chains are contractible if one of the timelike boundaries matches the spacelike boundaries.
From a physical perspective, this is because it's impossible to pick per-qubit X-basis data measurement outcomes (no matter how noisy) that violate the global stabilizer product being checked.
So the data measurements act as an inviolable source of truth that the stabilizer measurements are compared to, and corrected against, via the definitions of detection events.

With all of that established, I can finally define the steps making up a stability experiment, with an understanding of why each step is the way it is:

\begin{enumerate}

\item Choose a surface code patch with the same boundary type all around its perimeter (so that there is a global product of stabilizers with a known value).
The larger the patch you pick, the harder the experiment will be (since the number of stabilizers that are in the product, and have to  be correct, is larger for larger patches).

\item Initialize all data qubits into a basis that disagrees with the chosen boundary type (so that the relevant stabilizers are initially random, and aren't trivially forced to be correct by the error correction process).

\item Repeatedly measure the stabilizers of the surface code patch.
The number of repetitions is the code distance of the experiment, and can be anything you want.
The more times you repeat the stabilizer measurements, the more physical errors are needed to cause a logical error, and the more likely it is for the experiment to succeed (unless your physical error rate is over threshold).
(As with memory experiments, don't cheat by skipping measuring stabilizers in the basis that's technically not relevant to the final result.)

\item Measure all data qubits in a basis that disagrees with the chosen boundary type (so that the measurements can't be used as a source of truth that forces the relevant stabilizers to be correct via the error correction process).

\item Arbitrarily pick a round from the circuit's execution, and compute the product of all measurements from that round which correspond to stabilizers whose type matches the spatial boundaries.
Use a decoder to predict, based on the detection event data collected throughout the circuit's execution, if this product is flipped.
If the decoder predicts incorrectly (e.g. it says the product was flipped but the product was $+1$ as it would be in the noiseless case), a logical error has occurred.

\end{enumerate}

Those are the key ingredients of a stability experiment.
Having established this definition, let's now move on to quantifying how difficult the experiment is via simulation.

\section{Simulation Results}

When I originally came across the idea of a stability experiment, the motivating problem was to find something that would distinguish between different physical error mechanisms in the surface code.
Specifically, was there some way to make a high level error correction experiment that was more sensitive to measurement error than to data error (compared to a memory experiment).
I figured that because logical errors in the stability experiment look like repeated measurement errors, and because measurement errors can't cause logical errors in memory experiments (except by creating confusion that hides the key data errors), the stability experiment would be more sensitive to measurement errors.

To test this idea, I did a simulation of various memory experiments and stability experiments where I separately varied the amount of measurement noise and other noise.
In these experiments, syndrome decoding was done using correlated minimum weight perfect matching~\cite{fowler2013optimal}.
I simulated memory experiments with a fixed number of rounds and did a line fit of log logical error rate vs diameter (the code distance for memory experiments).
I used patch sizes 3x3, 5x5, and 7x7.
I also simulated stability experiments with a fixed patch size and did a line fit of log logical error rate vs rounds (the code distance for stability experiments).
I used 5, 15, and 25 rounds.
(I used the common surface code circuit schedule for surface code memory experiments, where the X and Z stabilizers are measured using four layers of CZs and hook errors are oriented to maximize the code distance \cite{horsman2012latticesurgery}.
I used the same schedule for the stability experiments, though the orientation of the hook errors should not matter there.)

\begin{figure}
    \centering
    \resizebox{\linewidth}{!}{
        \includegraphics{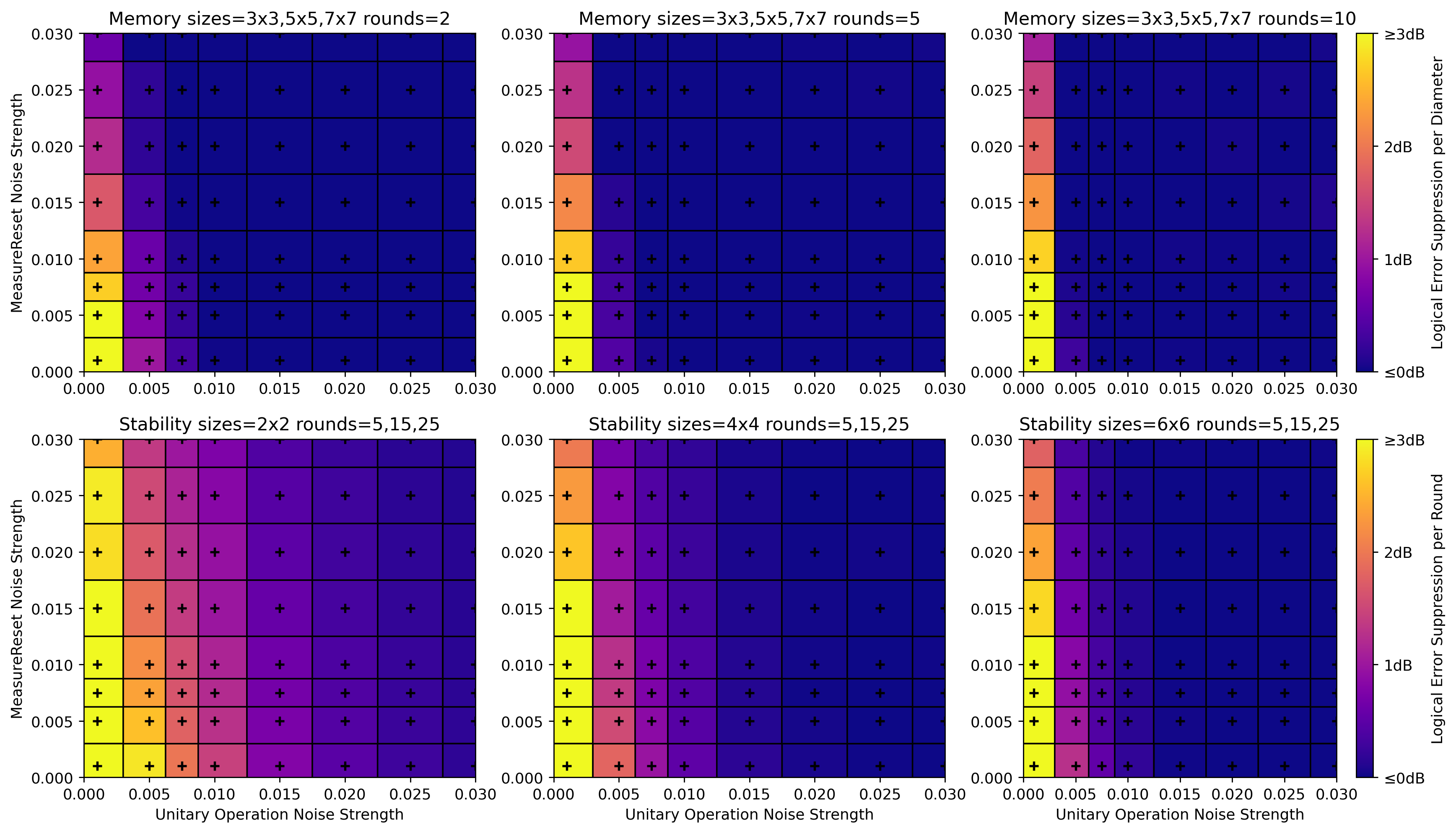}
    }
    \caption{
        Error suppression versus measure reset noise strength and unitary operation noise strength in memory experiments and stability experiments.
        A suppression factor of $3\text{dB}$ means there is a three decibel reduction (roughly a factor of two) in the logical error rate each time the code distance is increased by 1.
        When the suppression factor is $0\text{dB}$ or less, the logical error rate is not improving (or getting worse) as the code distance is increased.
        The suppression factor is derived from the slope of a line fit of log logical error rate versus code distance.
        For memory experiments the code distance is the patch diameter (the width or the height).
        For stability experiments the code distance is the number of rounds.
    }
    \label{fig:stability_sim_mr}
\end{figure}

I plotted a heatmap of the slopes of these lines; a heat map of how much the logical error rate was improving as the code distance was increased (see \fig{stability_sim_mr}).
This is a measure of how far (and whether) the physical noise rate is below the effective threshold of the experiment.

It turns out the stability experiment is not differently sensitive to measurement error.
Or, at least, not differently sensitive enough that the difference jumps out when looking at how much logical error rate improves as distance is increased.
Instead, what jumps out is that the stability experiment is overall much more forgiving than the memory experiment.
Its finite size thresholds are much higher.
For example, you can get exponential suppression of logical errors in 4x4 stability experiments using physical noise rates that are roughly twice as high as what's needed to get exponential suppression of logical errors in two-round memory experiments.
And the threshold of two-round memory experiments is already notably looser than the thresholds of longer memory experiments and the threshold of the surface code in fully general situations.

\begin{figure}
    \centering
    \resizebox{\linewidth}{!}{
        \includegraphics{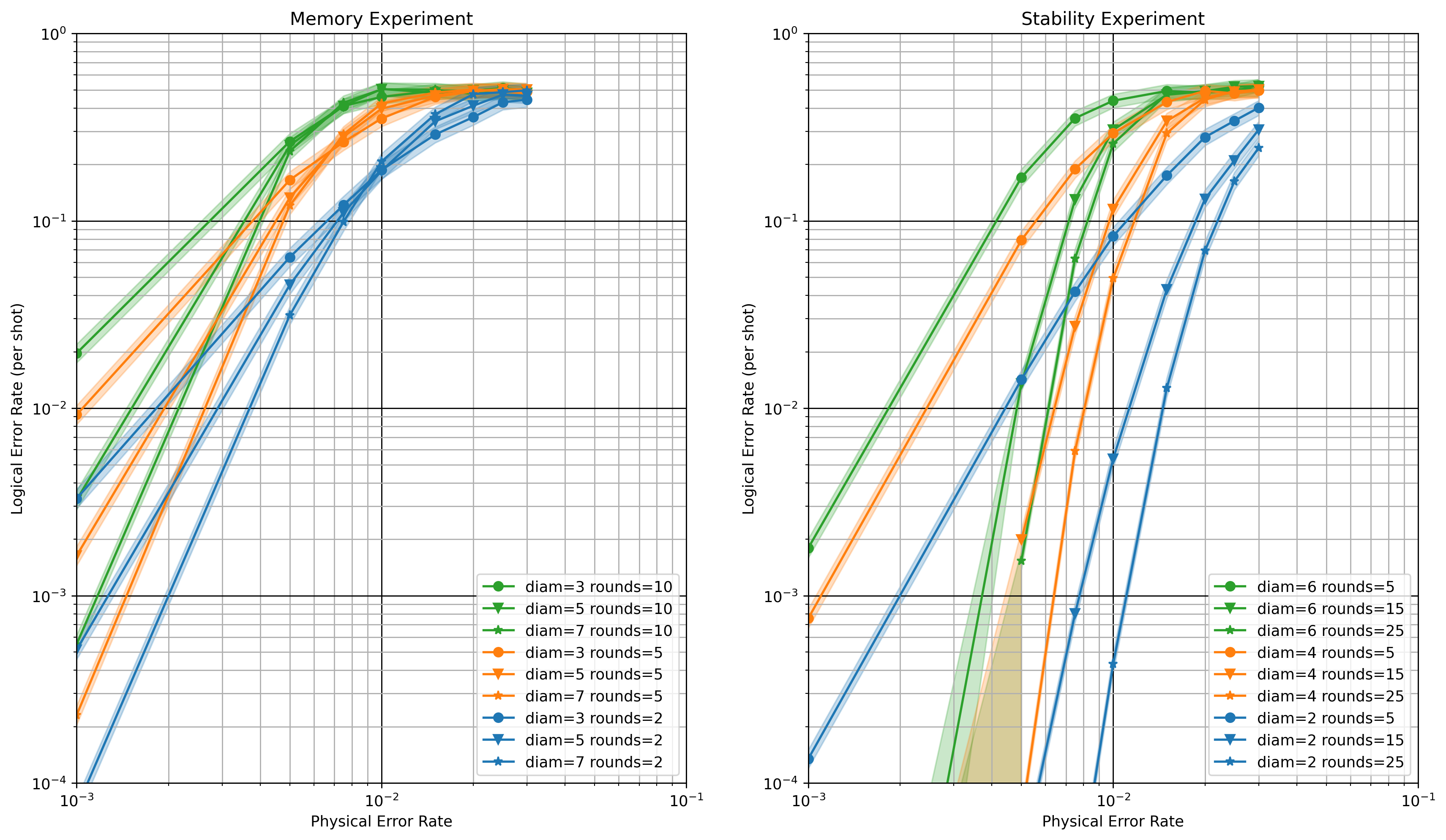}
    }
    \caption{
        Logical error rate versus physical error rate for memory and stability experiments.
        Uses the same data as \fig{stability_sim_mr}, but focuses on the case where the measure reset noise strength is equal to the depolarizing noise strength.
        Because it's easy to increase the code distance of the stability experiment (run more rounds, as opposed to build more qubits), the stability experiment curves are at higher code distances (5,15,25 instead of 3,5,7).
        This amplifies the benefits of being below threshold.
        Additionally, the stability experiment curves cross further to the right than the memory experiment curves, indicating that at small sizes the stability experiment has a more forgiving effective threshold.
        Color highlighting covers hypothesis logical error rates with a Bayes factor of less than 1000 relative to the maximum likelihood hypothesis (the plotted data point), given the sampled data.
    }
    \label{fig:stability_sim_threshold}
\end{figure}

In addition to plotting out the two dimensional heatmap, I did a more traditional threshold plot for the case with symmetric noise (see \fig{stability_sim_threshold}).
This plot even more strongly emphasizes the fact that the stability experiment, at small fixed patch sizes, is easier than memory experiments.

\section{Conclusion}

In this paper I described "stability experiments", which create a degenerate system that has a global spacelike invariant and then verify the invariant.
I argued that, although at first glance the stability experiment looks like it's not testing anything, it can directly measure relevant operational quantities, like how many times you need to repeat stabilizer measurements to hit a target logical error rate when moving a logical qubit.
Furthermore, I presented simulated data showing that stability experiments provide experimental access to the high code distance regime, with a forgiving threshold, at small patch sizes.

One of the underlying reasons that the stability experiment is ``easy" is that it is increasing only one dimension of the experiment: the one that increases the code distance.
In actual quantum computations, increasing protection against all error mechanisms requires growing the size of the computation in all three dimensions (two space, one time).
For example, consider moving a logical qubit.
The stability experiment corresponds to protecting one of the observables being moved.
To simultaneously increase the protection of the other observable, the surface code patch has to be widened.
To travel longer distances, the patch has to be lengthened.
Widening and lengthening the patch increases the number of stabilizers in the product, making the experiment harder, requiring lower physical error rates to achieve the same logical error rate.
This just-one-direction feature makes the stability experiment somewhat analogous to a repetition code.
Repetition codes also focus on just one error mechanism, and increase the dimension that increases protection against that error mechanism, while ignoring a second error mechanism that would be relevant in real world quantum computations.
Despite this dissimilarity to practice, repetition codes and stability experiments are useful for their simplicity and their ability to act as early warning systems for problems.
For example, in \cite{gqai2022suppressing}, a repetition code was used to detect a noise floor being caused by high energy events (such as cosmic rays).

\begin{figure}
    \centering
    \resizebox{0.4\linewidth}{!}{
        \includegraphics{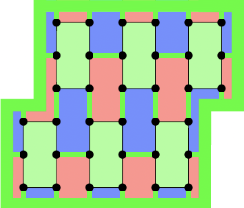}
    }
    \caption{
        Stabilizer configuration for a stability experiment in the planar honeycomb code \cite{hastings2021dynamically,gidney2021honeycombmemory,haah2021boundaries,gidney2022benchmarking,paetznick2022performance}.
        Uses RGB = XYZ coloring.
        Each edge is a parity measurement and each face is a parity check, with the qubits at the corners of the faces and with color indicating basis.
        The global spacelike invariant is that the product of all the green edges (Y basis parity measurements) should be equal to the product of all the green faces (Y basis parity checks formed from the X and Z edge measurements).
        This requires the entire boundary to have the same type, so that all boundary edges are measured in the same step.
    }
    \label{fig:honeycomb_stability}
\end{figure}

\begin{figure}
    \centering
    \resizebox{0.35\linewidth}{!}{
        \includegraphics{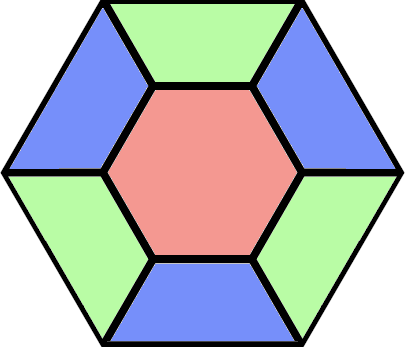}
    }
    \caption{
        Stabilizer configuration for a stability experiment in the color code \cite{bombin2006topological}.
        The coloring is just a three coloring, with no relationship to Pauli basis.
        Vertices are qubits.
        Each face represents both an $X$ type stabilizer to measure and a $Z$ type stabilizer to measure.
        There are two global spacelike invariants.
        The product of all blue regions should be equal to the product of all green regions, in both the X basis and the Z basis.
        This requires the boundary faces to alternate between the same two colors the entire way around.
    }
    \label{fig:color_code_stability}
\end{figure}

Although this paper has focused on doing the stability experiment in the surface code, the experiment generalizes to other error correcting codes.
\fig{honeycomb_stability} shows a stability experiment in the honeycomb code.
\fig{color_code_stability} shows a stability experiment in the color code.

\begin{figure}
    \centering
    \resizebox{\linewidth}{!}{
        \includegraphics{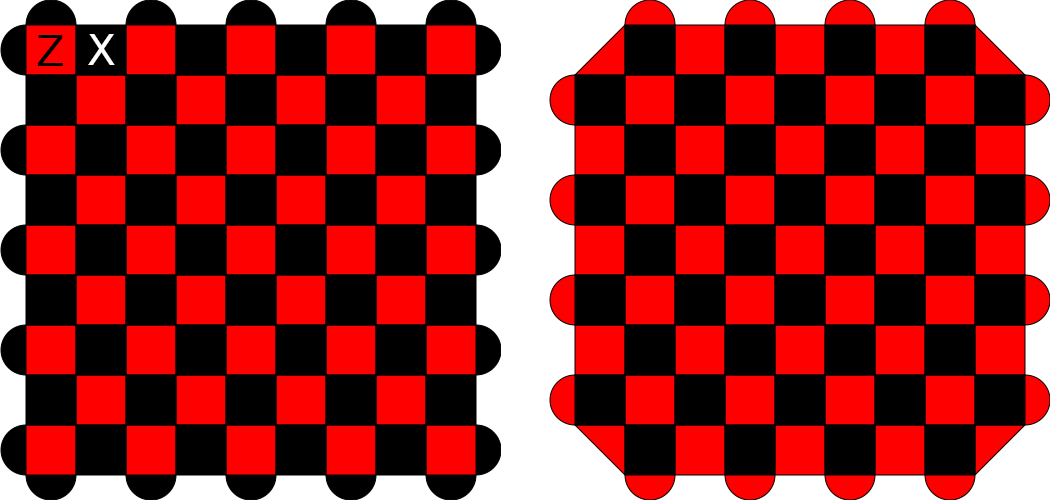}
    }
    \caption{
        Two 10x10 surface code patches, one with X type boundaries all the way around and one with Z type boundaries all the way around (and clipped corners).
        These stabilizer configurations can be used for stability experiments.
        The product of all 60 X stabilizers in the left patch should be +1.
        The product of all 57 Z stabilizers in the right patch should be +1.
    }
    \label{fig:10x10_XZ}
\end{figure}

Overall, I see no reason that stability experiments shouldn't be a standard benchmark for error correction systems.
I look forward to groups publishing how reliably and efficiently their machines can nail down ever larger stabilizer products (see \fig{10x10_XZ}).

\section{Acknowledgements}

Thanks to various other researchers at Google Quantum AI for useful feedback which influenced the paper, and apologies for not tracking exactly who gave feedback.
Thanks to Austin Fowler for writing the decoder that was used by this work.
Thanks to Hartmut Neven for creating an environment where this work was possible.

\printbibliography

\clearpage

\appendix

\clearpage
\section{Noise Model}
\label{app:noise_model}

\begin{table}[h!]
    \centering
    \begin{tabular}{|r|l|}
    \hline
    Ideal gate & Noisy gate
    \\
    \hline
    (Idling) & $\text{DEP1}(p_u)$
    \\
    (Single qubit gate) $U_1$ & $U_1 \circ \text{DEP1}(p_u)$
    \\
    $CZ$ & $CZ \circ \text{DEP2}(p_u)$
    \\
    \hline
    $R_Z$ & $R_Z \circ \text{XERR}(p_m)$
    \\
    $M_Z$ & $M_Z \circ \text{MERR}(p_m) \circ \text{DEP1}(p_m)$
    \\
    \hline
    \end{tabular}
    \caption{
        The noise model used by simulations in this paper.
        Gates in the left column are substituted for the noisy composition in the right column.
        The noise model has two parameters: $p_m$ (measure reset noise strength) and $p_u$ (unitary operation noise strength).
        Note $A \circ B = B \cdot A$ meaning $B$ is applied after $A$.
        Noise channels are defined in \tbl{noise_channels}.
    }
    \label{tbl:noise_model}
\end{table}

\begin{table}[h]
    \centering
    \begin{tabular}{|r|l|}
    \hline
    Noise channel & Probability distribution of effects
    \\
    \hline
    $\text{MERR}(p)$ & $\begin{aligned}
        1-p &\rightarrow \text{(report previous measurement correctly)}
        \\
        p &\rightarrow \text{(report previous measurement incorrectly; flip its result)}
    \end{aligned}$
    \\
    \hline
    $\text{XERR}(p)$ & $\begin{aligned}
        1-p &\rightarrow I
        \\
        p &\rightarrow X
    \end{aligned}$
    \\
    \hline
    $\text{ZERR}(p)$ & $\begin{aligned}
        1-p &\rightarrow I
        \\
        p &\rightarrow Z
    \end{aligned}$
    \\
    \hline
    $\text{DEP1}(p)$ & $\begin{aligned}
        1-p &\rightarrow I
        \\
        p/3 &\rightarrow X
        \\
        p/3 &\rightarrow Y
        \\
        p/3 &\rightarrow Z
    \end{aligned}$
    \\
    \hline
    $\text{DEP2}(p)$ & $\begin{aligned}
        1-p &\rightarrow I \otimes I
        &\;\;
        p/15 &\rightarrow I \otimes X
        &\;\;
        p/15 &\rightarrow I \otimes Y
        &\;\;
        p/15 &\rightarrow I \otimes Z
        \\
        p/15 &\rightarrow X \otimes I
        &\;\;
        p/15 &\rightarrow X \otimes X
        &\;\;
        p/15 &\rightarrow X \otimes Y
        &\;\;
        p/15 &\rightarrow X \otimes Z
        \\
        p/15 &\rightarrow Y \otimes I
        &\;\;
        p/15 &\rightarrow Y \otimes X
        &\;\;
        p/15 &\rightarrow Y \otimes Y
        &\;\;
        p/15 &\rightarrow Y \otimes Z
        \\
        p/15 &\rightarrow Z \otimes I
        &\;\;
        p/15 &\rightarrow Z \otimes X
        &\;\;
        p/15 &\rightarrow Z \otimes Y
        &\;\;
        p/15 &\rightarrow Z \otimes Z
    \end{aligned}$
    \\
    \hline
    \end{tabular}
    \caption{
        Definitions of various noise channels.
        Used by \tbl{noise_model}.
    }
    \label{tbl:noise_channels}
\end{table}

\clearpage
\section{Example Stability Experiment Circuit}

This is the stim circuit~\cite{stimcircuitformat}, without noise, from ``\path{b=Z,d=4,pd=0.001,pm=0.001,r=25,type=stability.stim}" in ``\path{circuits.zip}" which can be downloaded from the zenodo data set~\cite{craig_gidney_2022_6859486}.

\begin{lstlisting}[style=stimcircuit]
QUBIT_COORDS(0, 0) 0
QUBIT_COORDS(0, 1) 1
QUBIT_COORDS(0, 2) 2
QUBIT_COORDS(0, 3) 3
QUBIT_COORDS(1, 0) 4
QUBIT_COORDS(1, 1) 5
QUBIT_COORDS(1, 2) 6
QUBIT_COORDS(1, 3) 7
QUBIT_COORDS(2, 0) 8
QUBIT_COORDS(2, 1) 9
QUBIT_COORDS(2, 2) 10
QUBIT_COORDS(2, 3) 11
QUBIT_COORDS(3, 0) 12
QUBIT_COORDS(3, 1) 13
QUBIT_COORDS(3, 2) 14
QUBIT_COORDS(3, 3) 15
QUBIT_COORDS(-0.5, 0.5) 16
QUBIT_COORDS(-0.5, 2.5) 17
QUBIT_COORDS(0.5, -0.5) 18
QUBIT_COORDS(0.5, 0.5) 19
QUBIT_COORDS(0.5, 1.5) 20
QUBIT_COORDS(0.5, 2.5) 21
QUBIT_COORDS(0.5, 3.5) 22
QUBIT_COORDS(1.5, 0.5) 23
QUBIT_COORDS(1.5, 1.5) 24
QUBIT_COORDS(1.5, 2.5) 25
QUBIT_COORDS(2.5, -0.5) 26
QUBIT_COORDS(2.5, 0.5) 27
QUBIT_COORDS(2.5, 1.5) 28
QUBIT_COORDS(2.5, 2.5) 29
QUBIT_COORDS(2.5, 3.5) 30
QUBIT_COORDS(3.5, 0.5) 31
QUBIT_COORDS(3.5, 2.5) 32
R 0 1 2 3 4 5 6 7 8 9 10 11 12 13 14 15 16 17 18 19 20 21 22 23 24 25 26 27 28 29 30 31 32
TICK
H 1 3 4 6 7 9 11 12 13 14 15 16 17 18 19 20 21 22 23 24 25 26 27 28 29 30 31 32
TICK
CZ 0 19 1 20 2 21 3 22 4 23 5 24 6 25 8 27 9 28 10 29 11 30 12 31 14 32
TICK
H 0 1 2 3 5 6 8 9 10 11
TICK
CZ 0 16 2 17 1 19 5 20 3 21 7 22 8 23 6 24 10 25 9 27 13 28 11 29 15 30
TICK
H 1 3 4 11 12 14
TICK
CZ 0 18 4 19 2 20 6 21 5 23 9 24 7 25 8 26 12 27 10 28 14 29 13 31 15 32
TICK
H 4 5 6 7 9 10 12 13 14 15
TICK
CZ 1 16 3 17 4 18 5 19 6 20 7 21 9 23 10 24 11 25 12 26 13 27 14 28 15 29
TICK
H 5 7 10 13 15 16 17 18 19 20 21 22 23 24 25 26 27 28 29 30 31 32
TICK
M 16 17 18 19 20 21 22 23 24 25 26 27 28 29 30 31 32
DETECTOR(0.5, 0.5, 0) rec[-14]
DETECTOR(0.5, 2.5, 0) rec[-12]
DETECTOR(1.5, 1.5, 0) rec[-9]
DETECTOR(2.5, 0.5, 0) rec[-6]
DETECTOR(2.5, 2.5, 0) rec[-4]
SHIFT_COORDS(0, 0, 1)
TICK
REPEAT 23 {
    R 16 17 18 19 20 21 22 23 24 25 26 27 28 29 30 31 32
    TICK
    H 0 2 5 8 10 16 17 18 19 20 21 22 23 24 25 26 27 28 29 30 31 32
    TICK
    CZ 0 19 1 20 2 21 3 22 4 23 5 24 6 25 8 27 9 28 10 29 11 30 12 31 14 32
    TICK
    H 0 1 2 3 5 6 8 9 10 11
    TICK
    CZ 0 16 2 17 1 19 5 20 3 21 7 22 8 23 6 24 10 25 9 27 13 28 11 29 15 30
    TICK
    H 1 3 4 11 12 14
    TICK
    CZ 0 18 4 19 2 20 6 21 5 23 9 24 7 25 8 26 12 27 10 28 14 29 13 31 15 32
    TICK
    H 4 5 6 7 9 10 12 13 14 15
    TICK
    CZ 1 16 3 17 4 18 5 19 6 20 7 21 9 23 10 24 11 25 12 26 13 27 14 28 15 29
    TICK
    H 5 7 10 13 15 16 17 18 19 20 21 22 23 24 25 26 27 28 29 30 31 32
    TICK
    M 16 17 18 19 20 21 22 23 24 25 26 27 28 29 30 31 32
    DETECTOR(-0.5, 0.5, 0) rec[-34] rec[-17]
    DETECTOR(-0.5, 2.5, 0) rec[-33] rec[-16]
    DETECTOR(0.5, -0.5, 0) rec[-32] rec[-15]
    DETECTOR(0.5, 0.5, 0) rec[-31] rec[-14]
    DETECTOR(0.5, 1.5, 0) rec[-30] rec[-13]
    DETECTOR(0.5, 2.5, 0) rec[-29] rec[-12]
    DETECTOR(0.5, 3.5, 0) rec[-28] rec[-11]
    DETECTOR(1.5, 0.5, 0) rec[-27] rec[-10]
    DETECTOR(1.5, 1.5, 0) rec[-26] rec[-9]
    DETECTOR(1.5, 2.5, 0) rec[-25] rec[-8]
    DETECTOR(2.5, -0.5, 0) rec[-24] rec[-7]
    DETECTOR(2.5, 0.5, 0) rec[-23] rec[-6]
    DETECTOR(2.5, 1.5, 0) rec[-22] rec[-5]
    DETECTOR(2.5, 2.5, 0) rec[-21] rec[-4]
    DETECTOR(2.5, 3.5, 0) rec[-20] rec[-3]
    DETECTOR(3.5, 0.5, 0) rec[-19] rec[-2]
    DETECTOR(3.5, 2.5, 0) rec[-18] rec[-1]
    SHIFT_COORDS(0, 0, 1)
    TICK
}
R 16 17 18 19 20 21 22 23 24 25 26 27 28 29 30 31 32
TICK
H 0 2 5 8 10 16 17 18 19 20 21 22 23 24 25 26 27 28 29 30 31 32
TICK
CZ 0 19 1 20 2 21 3 22 4 23 5 24 6 25 8 27 9 28 10 29 11 30 12 31 14 32
TICK
H 0 1 2 3 5 6 8 9 10 11
TICK
CZ 0 16 2 17 1 19 5 20 3 21 7 22 8 23 6 24 10 25 9 27 13 28 11 29 15 30
TICK
H 1 3 4 11 12 14
TICK
CZ 0 18 4 19 2 20 6 21 5 23 9 24 7 25 8 26 12 27 10 28 14 29 13 31 15 32
TICK
H 4 5 6 7 9 10 12 13 14 15
TICK
CZ 1 16 3 17 4 18 5 19 6 20 7 21 9 23 10 24 11 25 12 26 13 27 14 28 15 29
TICK
H 0 1 2 3 4 6 8 9 11 12 14 16 17 18 19 20 21 22 23 24 25 26 27 28 29 30 31 32
TICK
M 0 1 2 3 4 5 6 7 8 9 10 11 12 13 14 15 16 17 18 19 20 21 22 23 24 25 26 27 28 29 30 31 32
DETECTOR(-0.5, 0.5, 0) rec[-50] rec[-17]
DETECTOR(-0.5, 2.5, 0) rec[-49] rec[-16]
DETECTOR(0.5, -0.5, 0) rec[-48] rec[-15]
DETECTOR(0.5, 0.5, 0) rec[-47] rec[-14]
DETECTOR(0.5, 1.5, 0) rec[-46] rec[-13]
DETECTOR(0.5, 2.5, 0) rec[-45] rec[-12]
DETECTOR(0.5, 3.5, 0) rec[-44] rec[-11]
DETECTOR(1.5, 0.5, 0) rec[-43] rec[-10]
DETECTOR(1.5, 1.5, 0) rec[-42] rec[-9]
DETECTOR(1.5, 2.5, 0) rec[-41] rec[-8]
DETECTOR(2.5, -0.5, 0) rec[-40] rec[-7]
DETECTOR(2.5, 0.5, 0) rec[-39] rec[-6]
DETECTOR(2.5, 1.5, 0) rec[-38] rec[-5]
DETECTOR(2.5, 2.5, 0) rec[-37] rec[-4]
DETECTOR(2.5, 3.5, 0) rec[-36] rec[-3]
DETECTOR(3.5, 0.5, 0) rec[-35] rec[-2]
DETECTOR(3.5, 2.5, 0) rec[-34] rec[-1]
SHIFT_COORDS(0, 0, 1)
DETECTOR(0.5, 0.5, 0) rec[-33] rec[-32] rec[-29] rec[-28] rec[-14]
DETECTOR(0.5, 2.5, 0) rec[-31] rec[-30] rec[-27] rec[-26] rec[-12]
DETECTOR(1.5, 1.5, 0) rec[-28] rec[-27] rec[-24] rec[-23] rec[-9]
DETECTOR(2.5, 0.5, 0) rec[-25] rec[-24] rec[-21] rec[-20] rec[-6]
DETECTOR(2.5, 2.5, 0) rec[-23] rec[-22] rec[-19] rec[-18] rec[-4]
OBSERVABLE_INCLUDE(0) rec[-17] rec[-16] rec[-15] rec[-13] rec[-11] rec[-10] rec[-8] rec[-7] rec[-5] rec[-3] rec[-2] rec[-1]
\end{lstlisting}

\end{document}